\documentclass[12pt]{article}

\usepackage{amsmath}
\usepackage{amssymb}
\usepackage{euscript}

\usepackage{epsfig}

\usepackage{cite}

\usepackage{alltt}

\def\Order#1{${\cal O}(#1$)}

\def\lambdag{{\lambda_{\gamma}}}
\def\lambdag{{\lambda}}
\def\draftdate{\relax}
\def\mda{\relax}
\def\mua{\relax}
\def\mla{\relax}
\def\draft{
\def\thtystars{******************************}
\def\sixtystars{\thtystars\thtystars}
\typeout{}
\typeout{\sixtystars**}
\typeout{* Draft mode!
         For final version remove \protect\draft\space in source file *}
\typeout{\sixtystars**}
\typeout{}
\def\draftdate{\today}
\def\mua{\marginpar[\boldmath\hfil$\uparrow$]%
                   {\boldmath$\uparrow$\hfil}%
                    \typeout{marginpar: $\uparrow$}\ignorespaces}
\def\mda{\marginpar[\boldmath\hfil$\downarrow$]%
                   {\boldmath$\downarrow$\hfil}%
                    \typeout{marginpar: $\downarrow$}\ignorespaces}
\def\mla{\marginpar[\boldmath\hfil$\rightarrow$]%
                   {\boldmath$\leftarrow $\hfil}%
                    \typeout{marginpar: $\leftrightarrow$}\ignorespaces}
\overfullrule 5pt
\oddsidemargin -5mm
\marginparwidth 29mm
}

\begin{document}                     

\allowdisplaybreaks

\begin{titlepage}

\begin{center}
\end{center}
\vspace{-28mm}

\begin{flushright}
{\bf  CERN-TH/2001-274\\
      DESY 01-184\\
      LAPP-EXP 2002-01\\
      MPI-PhT-2001-40\\
      PSI-PR-02-01\\
      UTHEP-01--1001
}
\end{flushright}

\vspace{1mm}
\begin{center}
{\Large\bf
On Theoretical Uncertainties of the \boldmath{$W$} Angular \\[0.4ex]
\! Distribution
in \boldmath{$W$}-Pair Production at LEP2 Energies$^{\star}$\!
}
\end{center}

\vspace{1mm}
\begin{center}
{\bf R.~Bruneli\`ere$^{a}$, A.~Denner$^b$, S.~Dittmaier$^{c\dagger}$,  
  S.~Jadach$^{d,e}$, S.~J\'ez\'equel$^{a}$, W.~P\l{}aczek$^{f,e}$, 
  M.~Roth$^g$, M.~Skrzypek$^{d,e}$, D.~Wackeroth$^h$,  
  B.F.L.~Ward$^{i,j,k}$} {\em and}~{\bf Z.~W\c{a}s$^{d,e}$}

\vspace{1mm}
{\em
$^a$ LAPP, IN2P3-CNRS, Chemin de Bellevue, F-74940 Annecy-Le-Vieux, France\\
$^b${Paul Scherrer Institut, CH-5232 Villigen PSI, Switzerland}\\
$^c${Deutsches Elektronen-Synchrotron DESY, D-22603 Hamburg, Germany}\\
$^d$Institute of Nuclear Physics,
    ul. Kawiory 26a, 30-055 Cracow, Poland\\
$^e$CERN, TH Division, CH-1211 Geneva 23, Switzerland\\
$^f$Institute of Computer Science, Jagellonian University,\\
   ul. Nawojki 11, 30-072 Cracow, Poland\\
$^g${Institut f\"ur Theoretische Physik, Universit\"at Karlsruhe
    D-76131 Karlsruhe, Germany}\\
$^h${Department of Physics and Astronomy, University of
     Rochester,\\  Rochester, NY 14627-0171, USA}\\
$^i$Max-Planck-Institut f\"ur Physik, 80805 Munich, Germany\\
$^j$Department of Physics and Astronomy,\\
   The University of Tennessee, Knoxville, TN 37996-1200, USA\\
$^k$SLAC, Stanford University, Stanford, CA 94309, USA
}
\end{center}

\begin{abstract}
  We discuss theoretical uncertainties of the distribution in
  the cosine of the $W$ polar
  angle projected into a measurement of the anomalous triple
  gauge-boson coupling
  $\lambda=\lambda_{\gamma}=\lambda_Z$
  at LEP2 energies for the tandem of the Monte Carlo event generators
  KoralW and YFSWW3 and for the Monte Carlo event generator RacoonWW.
  Exploiting numerical results of these programs and cross-checks with
  experimental fitting procedures, we estimate that the theoretical
  uncertainty of the value of $\lambda$
  due to electroweak corrections, as obtained at LEP2 with the help of
  these programs, is $\sim 0.005$, about half of the expected
  experimental error for the combined LEP2 experiments ($\sim 0.010$).
  We use certain idealized event selections; 
  however, we argue that
  these results are valid for realistic LEP2 measurements.
\end{abstract}

\begin{center}
{\it To be submitted to Physics Letters B}
\end{center}

\footnoterule
\noindent
{\footnotesize
\begin{itemize}
\item[${}^{\star}$]
  Work partly supported by 
  the Polish Government grants KBN 5P03B12420 
  and KBN 5P03B09320, 
  the European Commission 5th framework contract HPRN-CT-2000-00149,
  the  Swiss Bundesamt f\"ur Bildung und Wissenschaft contract 99.0043, 
  Polish-French Collaboration within IN2P3 through LAPP Annecy,
  the US DOE Contracts DE-FG02-91ER40685, DE-FG05-91ER40627 and
  DE-AC03-76ER00515.
\item[${}^{\dagger}$]
Heisenberg Fellow of the Deutsche Forschungsgemeinschaft.
\end{itemize}
}

\vspace{-1mm}
\begin{flushleft}
{\bf 
January 2002
}
\end{flushleft}

\end{titlepage}

A distribution of the {\em cosine} of the $W$ boson production angle
$\theta_W$ is one of the main observables in the $W$-pair production
process measured by the LEP2 experiments. It is sensitive to the
triple gauge-boson
couplings (TGCs) $WWV$, with $V=Z,\gamma$
\cite{Aleph-TGC:2001,Delphi-TGC:2001,L3-TGC:1999,Opal-TGC:2000,lepwwac},
particularly to the $C$- and $P$-conserving anomalous couplings
$\lambda_{\gamma},\lambda_Z$~\cite{TGC-YR:1996}.  In this work we
present our estimate of the theoretical uncertainties (TU) related to
electroweak (EW) corrections in the measurement of the anomalous TGC
$\lambda=\lambda_{\gamma}=\lambda_Z$ in the $W$-pair production at
LEP2 energies%
\footnote{A similar analysis for the $W$ mass measurement was done in
          ref.~\cite{Jadach:2001cz}.
          }.
It is important to note that we treat the anomalous TGC as a small new-physics
effect beyond the consistent Standard-Model prediction, i.e.\ we
introduce the anomalous TGC in the lowest-order matrix element, which
is dressed by initial-state radiation, while genuine weak corrections are
unaffected by the anomalous TGC.
We concentrate on $\lambdag$ in order to keep our analysis as simple
as possible and because the value of $\lambdag$ fitted to experimental data
depends more strongly on the shape of the $\theta_W$ distribution than on
its total normalization. 
In fact, while $\lambdag$ is mainly sensitive to the shape of the $W$ polar 
angle distribution, it is not particularly sensitive to other single
distributions. Using only this variable, we lose only 30\% in 
sensitivity~\cite{TGC-YR:1996}. On the other hand, the analysis
of other TGCs would require using more observables.

Our results were obtained using the Monte Carlo (MC) event generators
YFSWW3~\cite{yfsww2:1996,yfsww3:1998,yfsww3:1998b,yfsww3:2000a,yfsww3:2001}
and KoralW~\cite{koralw:1995a,koralw:1998,koralw:2001}, as well as
RacoonWW~\cite{Denner:1999gp,Denner:1999kn,Denner:2000bj,Denner:2001zp,Denner:2001vr}.
In YFSWW3 and KoralW, the anomalous TGCs are implemented according to the most
general parametrization of ref.~\cite{hagiwara:1987}, and also
according to two simplified parametrizations given
in ref.~\cite{TGC-YR:1996}.
In RacoonWW%
\footnote{The anomalous TGCs are implemented in RacoonWW~1.2, which was
  released recently 
  (see {\tt http://ltpth.web.psi.ch/racoonww/racoonww.html}).},
the anomalous TGCs involving $W$ bosons are implemented according to the 
most general parametrization of ref.~\cite{hagiwara:1987} in the form given
in ref.~\cite{TGC-YR:1996}, while those involving only neutral gauge bosons
are implemented according to refs.~\cite{Gounaris:1999kf,Gounaris:2000dn}.

In the following, we consider the semi-leptonic process $e^+ e^-
\rightarrow u\bar{d} \mu^-\bar{\nu}_{\mu}$, which belongs to the
so-called CC11 class. The corresponding Feynman diagrams constitute a
gauge-invariant subset of all 4-fermion final-state processes (see
e.g.\ ref.~\cite{LEP2YR:1996} for more details).  We shall study only
the leptonic $W$ polar angle, i.e.\ the one reconstructed from the
four-momenta of the $\mu^-$ and $\bar{\nu}_{\mu}$ (in the actual
experiments, the neutrino four-momentum is reconstructed from the
constrained kinematical fit, see e.g.\ 
refs.~\cite{Aleph-TGC:2001,Delphi-TGC:2001,L3-TGC:1999,Opal-TGC:2000}).
The input parameters are the same as in the LEP2 MC Workshop studies
performed in 2000~\cite{4f-LEP2YR:2000}.  All the results in this work
are given for the centre-of-mass energy $E_{CM} = 200\,$GeV.  The
results from YFSWW3 are for the leading-pole approximation (LPA) for
the double-resonant $WW$ production and decay of the type advocated in
ref.~\cite{stuart:1997}, which is implemented in YFSWW3 as an option
LPA$_a$ (see ref.~\cite{yfsww3:2001} for more details).  The results
from RacoonWW are based on the double-pole approximation (DPA)
as worked out in ref.~\cite{Denner:2000bj}.  In all our numerical
exercises we use $\lambda=\lambda_{\gamma}=\lambda_Z$, and all other
anomalous couplings are set to zero.

In our analysis we use two different fitting procedures. 
In the first part of the paper we do direct fits to the $\cos\theta_W$
distributions with the help of the semi-analytical program 
KorWan~\cite{koralw:1995a,koralw:1995b,koralw:1998}. 
Since in KorWan one cannot apply any experimental-like cuts,
it can be used, in principle, only for fitting some idealistic 
distributions (without cuts), for which 
good fits (with low $\chi^2$) can be expected. 
We, however, decided to apply the KorWan fits also
to more experimental-like distributions. 
The results of these fits
should be considered only as an independent cross-check for the main
fitting procedure, the so-called Monte Carlo parametric fit, presented
in the second part of the paper. In this procedure, contrary to the
KorWan fits, any event selection criteria can be applied.
As such, this fitting procedure can be used for experimental data,
which is not the case for the KorWan-based method.
Note that the results of the KorWan fits depend on the Monte Carlo
errors, and thus on the models used for the Monte Carlo integration.
Since the models implemented in KorWan and YFSWW3 are similar, fits of
the YFSWW3 distributions yield reasonable results for the shifts in
$\lambdag$.  However, since the models implemented in KorWan and
RacoonWW are too different to make reasonable fits possible, we do not
fit the RacoonWW distributions in the first part of the paper.

\begin{figure}[!ht]
\centering
\setlength{\unitlength}{0.1mm}
\epsfig{file=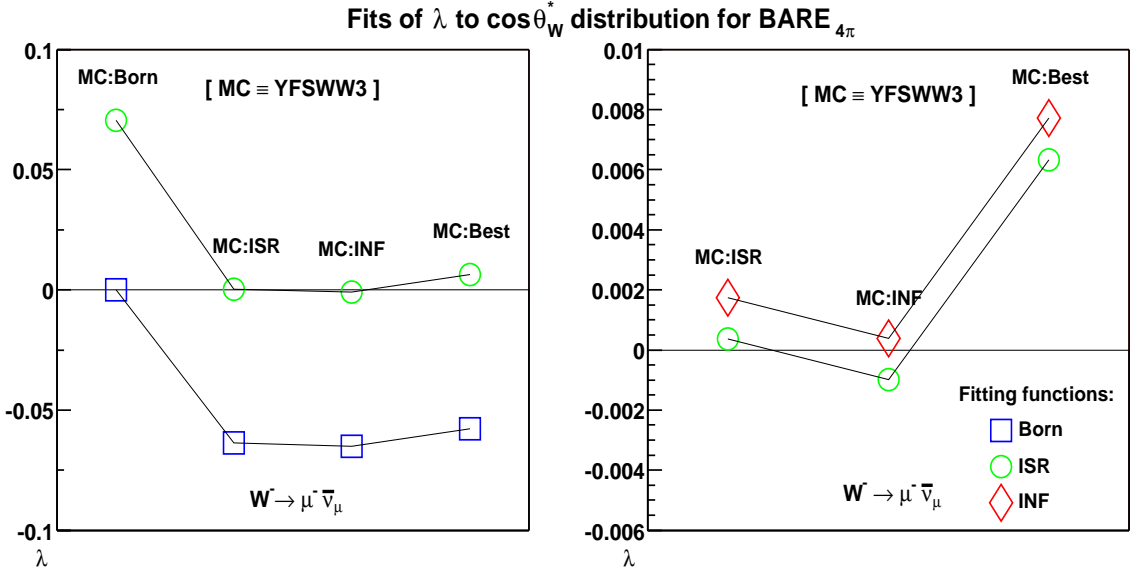,width=160mm,height=80mm}
\caption{\small\sf
  Introductory exercise with YFSWW3, see description in text.
}
\label{fig:Fig1}
\end{figure}
In the first preparatory step, we construct a simple fitting procedure
for $\cos\theta_W^*$, where the superscript $*$ means that $\theta_W$
is evaluated in the $WW$ rest frame%
\footnote{%
More precisely, $\theta_W^*$ is defined as the angle between the
directions of the $W^-$ in the $WW$ rest frame and the $e^-$ beam in
the laboratory frame.}. 
We ``calibrate'' the fitting procedure for the $\cos\theta_W^*$
distribution by using it with the MC data in which we switch 
on/off the same effects as in the fitting function (FF), typically the
initial-state radiation (ISR) and the non-factorizable (NF) corrections,
in order to see whether we get agreement in the case of the same
effect in the MC data and in the FF.  The FF is taken, in all cases,
from the semi-analytical program
KorWan~\cite{koralw:1995a,koralw:1995b,koralw:1998}%
\footnote{The relevant distribution will be available in the next
  release of KorWan/KoralW.}.  In the YFSWW3 MC and in the fitting
function the NF corrections are implemented in the inclusive
approximation (denoted by INF in the following) of the so-called
screened Coulomb ansatz by Chapovsky and
Khoze~\cite{Chapovsky:1999kv}, which is an approximation of the full
calculation of the non-factorizable
corrections~\cite{Beenakker:1997bp,Beenakker:1997ir,Denner:1998rh,Denner:1998ia}.
Here, for completeness, we note that in RacoonWW non-factorizable
corrections are included beyond the inclusive approximation (see
Ref.~\cite{Denner:2000bj}). In particular, the real-photonic
non-factorizable corrections (as well as final-state radiation) are
contained in the full $e^+e^-\to4f+\gamma$ matrix elements which are
used.

One immediate profit of this introductory exercise is that we get an
estimate of the size of the ISR and NF effects on the fitted
(measured) $\lambdag$ in Fig.~\ref{fig:Fig1}, which is determined from
a two-parameter fit to the $\cos\theta_W^*$ distribution. In addition
to $\lambdag$ we also fit the overall normalization, which is
necessary in the presence of cuts in the MC data and/or when the MC
model does not correspond exactly to the KorWan model. As a
consequence, the fitted values of $\lambdag$ depend on the
normalization and on the shape of the distribution.  The results of
the first exercise, for the BARE$_{4\pi}$ event selection (no photon
recombination) and without any cuts (the subscript $4\pi$ means the
full solid-angle coverage), are shown in Fig.~\ref{fig:Fig1}.  Let us
explain briefly the notation: {\tt Born} denotes the CC03 Born-level
results, {\tt ISR} the ones including the \Order{\alpha^3} LL YFS
exponentiation for the ISR as well as the standard Coulomb
correction~\cite{khoze:1995}, {\tt INF} the above plus the INF
correction, and {\tt Best} denotes the best predictions from YFSWW3,
i.e.\ all of the above plus the \Order{\alpha^1} EW non-leading (NL)
corrections%
\footnote{The \Order{\alpha^1} EW corrections for the $WW$ production
  stage in YFSWW3 are based on
  refs.~\cite{fleischer:1989,fleischer:1994}.}.  
In all the MC simulations the input value of $\lambdag$ was $0$.  The
errors of the fitted values due to the statistical errors of the MC
results are in all cases $<0.001$.

Let us summarize the observations resulting from Fig.~\ref{fig:Fig1}:
\begin{itemize}
\item
  The fitted values of $\lambdag$ agree with the input ones, i.e.\
  $\lambdag=0$,
  in the cases when the FF corresponds to the model used in the MC
  simulations.
\item
  The effects of the ISR are huge, $\sim 0.07$.
\item 
  The effects of the INF corrections are very small, $\sim 0.0015$.
This is to be expected since the non-factorizable corrections are
strongly suppressed once the invariant masses of the $W$~bosons are
integrated over.
\item
  The effects of the NL corrections are sizeable, $\sim 0.008$.
\end{itemize}

\begin{figure}[!ht]
\centering
\setlength{\unitlength}{0.1mm}
\begin{picture}(1600,1600)
\put(  0,800){\makebox(0,0)[lb]{
\epsfig{file=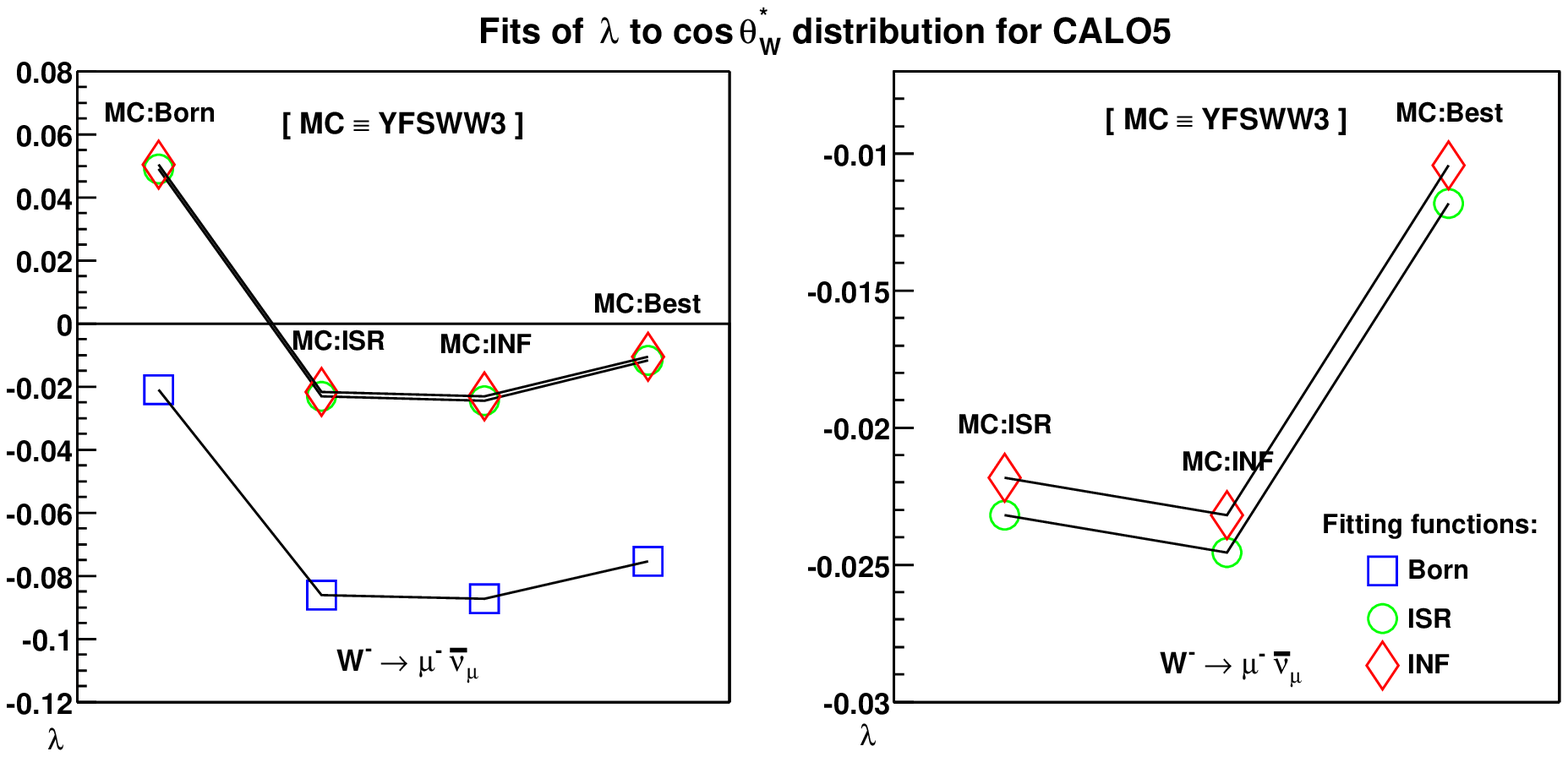,width=160mm,height=80mm}
}}

\put(  0,0){\makebox(0,0)[lb]{
\epsfig{file=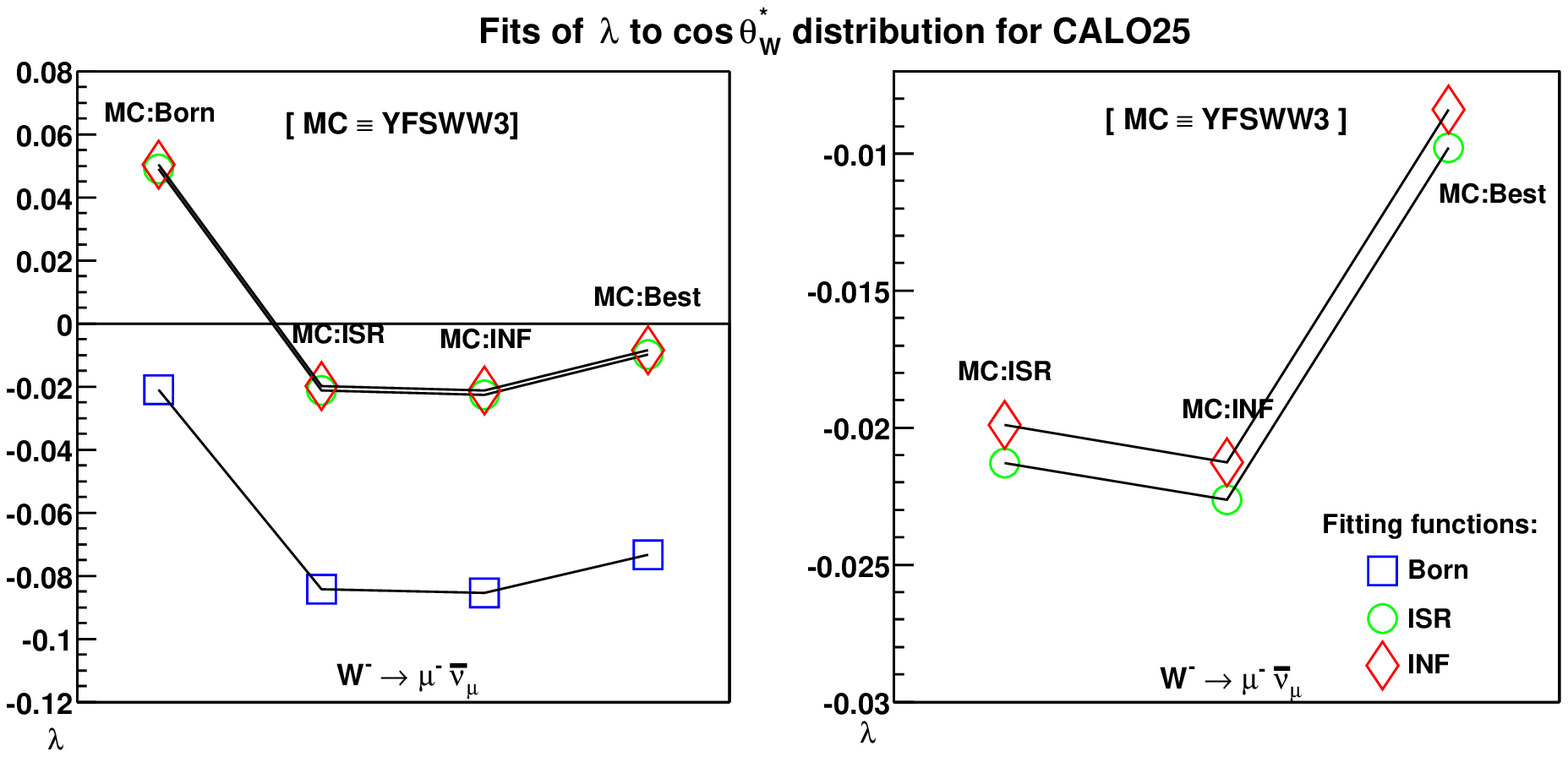,width=160mm,height=80mm}
}}
\end{picture}
\caption{\small\sf
  Results for calorimetric-type acceptances from YFSWW3.
}
\label{fig:Fig2}
\end{figure}

Before we go to the next exercise, let us briefly describe the
acceptances/cuts that were used in the MC simulations for the
following calculations:
\begin{enumerate}
\item
  All photons within a cone of $5^{\circ}$ around the beams were treated
  as {\em invisible}, 
  i.e.\ their momenta were discarded when calculating angles, energies,
  and invariant masses.
\item
  The invariant mass of a {\em visible} photon with each 
  charged final-state fermion, $M_{f_{ch}}$, was calculated, 
  and the minimum value 
  $M^{min}_{f_{ch}}$ was found. If $M^{min}_{f_{ch}}< M_{rec}$ 
  or if the photon energy $E_{\gamma}<1\,$GeV,
  the photon was combined with the corresponding fermion, 
  i.e.\ the photon four-momentum was added to the fermion four-momentum
  and the photon was discarded. 
  In YFSWW3 and KoralW,
  this was repeated for all {\em visible} photons,
  while in RacoonWW there is only one visible photon.

  In our numerical tests we used two values of the recombination cut: 
  \begin{displaymath}
  M_{rec} = \left\{
  \begin{tabular}{l}
   \: 5\,{\rm GeV:} \hspace{1cm} {\rm CALO5}, \\
   25\,{\rm GeV:} \hspace{1cm} {\rm CALO25}.
  \end{tabular}  
  \right.
  \end{displaymath}
  Let us remark that we have changed here the labelling of these
  recombination cuts from the slightly misleading {\em bare} and {\em calo}
  names used in Ref.~\cite{4f-LEP2YR:2000}. They correspond to our
  CALO5 and CALO25, respectively.
  This change allows us to reserve the BARE
  name for a ``truly bare final fermion'' event selection (without any recombination).
\item
  Finally, we required that the polar angle of any charged final-state
  fermion with respect to the beams be $\theta_{f_{ch}}>10^{\circ}$.
\end{enumerate}

In the next exercise, presented in Fig.~\ref{fig:Fig2}, we examine
similar effects as in Fig.~\ref{fig:Fig1}, but now for the calorimetric-type 
acceptances/cuts: CALO5 and CALO25. 
\\
The following observations resulting from Fig.~\ref{fig:Fig2} can be made:
\begin{itemize}
\item 
  For the calorimetric event selections, the fitted values of $\lambdag$
  are shifted by $\sim -0.02$ with respect to the input ones
  in the cases when the FF corresponds to the model used in the MC
  simulations.
  This can be explained by the fact that the FF of KorWan is for the
  full angular acceptance while the MC results were obtained with a cut
  on the charged final-state fermions.
\item
  The size of the ISR is the same as for the BARE event 
  selection, cf. Fig.~\ref{fig:Fig1}.
\item 
  The shift of $\lambdag$ due to the NL corrections increased
  from $\sim 0.008$ for BARE to $\sim 0.011$ for both CALO5 and CALO25.
\item 
  The fitted values of $\lambdag$ for CALO5 are
  slightly different (by $\sim 0.002$) from the corresponding ones for
  CALO25, but the differences $\Delta\lambdag$ between various models
  are the same.  
\end{itemize}

From the fits to KorWan, we can conclude that the effects of the NL 
corrections on $\lambdag$ are sizeable, of the order of the expected
experimental precision of the final LEP2 data analysis.
Thus, they need to be accounted for in the experimental measurements
of the TGCs. 

In the second, main 
part of our study, we follow an alternative fitting
strategy for $\lambdag$ in which we do not exploit the angular
distribution (fitting function) of KorWan, but instead we rely
entirely on the MC results,
more precisely, on a simple polynomial parametrization of the 
normalized angular distribution
$D(\cos\theta_W,\lambdag)=
\frac{1}{\sigma} \frac{d\sigma}{d\cos\theta_W}(\cos\theta_W,\lambdag)$ 
as a function of $\cos\theta_W$ and $\lambdag$, determined from running 
the MC for several values of $\lambdag$.  
Since the distribution $D$ is normalized, the fitted values of
$\lambdag$ depend only on the shape.
Let us call this method a ``Monte Carlo
parametric fit'', or the MPF in short, and the MPF fitting function --
the MPFF.  Once such a MPF with the MPFF is established, then in
principle it could be used to quantify a deviation of $\lambdag$ by
fitting the MPFF to the experimental angular distribution.  Our aim is
rather to quantify various components of TU of the SM predictions of
the MC tandem KoralW and YFSWW3 and the MC
RacoonWW directly in terms of $\lambdag$ (similarly as in the
previous method based on KorWan).  The main advantage of the MPF is
that this can be applied for an arbitrary event selection and any
definition of $\theta_W$, while any semi-analytical program like
KorWan has a strongly restricted choice of the angle definition and
kinematic cuts.  The MPF method is feasible and not very difficult in
practice because the distribution $D(\cos\theta_W,\lambdag)$ is a very
smooth function of its two arguments for LEP2 energies%
\footnote{The MPF procedure would be less practical at high energies
  or even at LEP2 for fitting $M_W$ using an effective $W$ mass
  distribution.}.   
Another advantage of the MPF will unfold in the following -- in fact,
we shall be able to get closer to the fitting procedure of the real
LEP2 experiments.

Here, we include also the results from RacoonWW.
The comparison of YFSWW3 and RacoonWW is very interesting, because
the two calculations differ almost in every aspect of the implementation
of the ISR, final-state radiation (FSR), NL, and NF corrections.
The ``ISR'' prediction of RacoonWW, which also contains the off-shell
Coulomb singularity \cite{Fadin:1993kg,Bardin:1993mc,Denner:1998ia},
is exclusively based on the collinear structure-function approach,
i.e.\ all generated photons are collinear to the beams and are thus
treated as invisible.  In contrast, YFSWW3 generates photons with
finite transverse momenta also in the ``ISR'' version. 
For the ``Best'' predictions of RacoonWW, photon
radiation in \Order{\alpha} is included via the full matrix elements,
and the main differences to YFSWW3 come from two sources. First,
RacoonWW does not contain ISR corrections to the non-collinear
one-photon emission contribution, in contrast to YFSWW3. Then, while
the FSR is generated by PHOTOS~\cite{photos:1994} in YFSWW3, it is
included via the explicit $e^+e^-\to4f+\gamma$ matrix elements in RacoonWW.

How is the MPF realized in practice?
We use the following 9-parameter MPFF formula
\begin{equation}
\rho(\cos\theta_W,\lambdag)=\frac{D(\cos\theta_W,\lambdag)}{D(\cos\theta_W,0)}
=\sum_{i=0}^2 a_i \lambdag^i +\cos\theta_W \sum_{i=0}^2 b_i 
\lambdag^i+\cos^2\theta_W \sum_{i=0}^2 c_i \lambdag^i,
\end{equation}
where parameters $a_i$, $b_i$ and $c_i$, $i=0,1,2$, are determined by fitting
the
$\rho(\cos\theta_W,\lambdag)$ distribution obtained from YFSWW3 
or RacoonWW
for the three values $\lambdag = -0.2,\, 0.0,\, 0.2\;$%
\footnote{The actual fit of $a_i$, $b_i$ and $c_i$ is done in two steps,
  first for each fixed $\lambdag = -0.2,\, 0.0,\, 0.2$, and next for
  the $\lambdag$ dependence. We have checked that the results do not change
  if the fit is done in one step.
}.
We have checked that the fitted MPFF: 
(a) reproduces the MC result of $\rho(\cos\theta_W,\lambdag)$
for any $\lambdag\in (-0.2,0.2)$ within the fit error of $\sim 0.001$,
and (b) when the MPFF is used to fit $\rho(\cos\theta_W,\lambdag)$
generated by the MC for any $\lambdag\in (-0.2,0.2)$
the fitted value agrees, within the fit error, with the input 
$\lambdag$ used in the MC.
Moreover, within fit errors we find $a_0=1$ and $b_0=c_0=0$ as
required.
Obviously, the other coefficients $a_i$, $b_i$ and $c_i$ are different
for every kind of the event selection and angle definition, and the MPF
is always a two-step procedure: first we determine the MPFF using the MC,
and then we apply it to the results of another MC run
(we do not fit experimental data).
All the results in the following are obtained for the $\cos\theta_W^{\rm LAB}$
distributions, i.e. from now on we use $\theta_W \equiv \theta_W^{\rm LAB}$.

\begin{table}[!ht]
{
\begin{center}
\begin{tabular}{||l|c||c|c|c|c||}
\hline\hline
\multicolumn{2}{||c||}{\bf Fitting procedure}  & 
\multicolumn{4}{|c||}{\bf Fitted data } \\
\hline
  Fitting function  & Accept. & {Y: Best$-$ISR}  
  & {R: Best$-$ISR}  & {Best: R$-$Y} & Accept.\\
\hline\hline
~1.  KorWan                & BARE   & $0.0114\,(6)$ & --- 
    & ---  & CALO5 \\
~2.  KorWan                & BARE   & $0.0115\,(6)$ & --- 
    & ---  & CALO25\\
\hline
~3. {\small MPF: Y-ISR}  & CALO5  & $0.0112\,(7)$ & $0.0097\,(8)$ 
    & $0.0008\,(9)~\,$  & CALO5 \\
~4. {\small MPF: Y-ISR}  & CALO5  & $0.0115\,(7)$ & $0.0161\,(8)$
    & $0.0008\,(9)~\,$  & CALO25\\
\hline
~5. {\small MPF: R-ISR}  & CALO5  & $0.0112\,(7)$ & $0.0097\,(8)$ 
    & $0.0008\,(9)~\,$  & CALO5 \\
~6. {\small MPF: R-ISR}  & CALO5  & $0.0115\,(7)$ & $0.0161\,(8)$ 
    & $0.0008\,(10)$  & CALO25\\
\hline
~7. {\small MPF: Y-Best} & CALO5  & $0.0113\,(7)$ & $0.0098\,(8)$ 
    & $0.0008\,(10)$  & CALO5 \\
~8. {\small MPF: Y-Best} & CALO5  & $0.0116\,(7)$ & $0.0162\,(8)$ 
    & $0.0008\,(9)~\,$  & CALO25\\
\hline
~9. {\small MPF: R-Best} & CALO5  & $0.0110\,(7)$ & $0.0096\,(8)$ 
    & $0.0007\,(9 )~\,$  & CALO5 \\
10. {\small MPF: R-Best} & CALO5  & $0.0113\,(7)$ & $0.0158\,(8)$ 
    & $0.0008\,(9 )~\,$  & CALO25\\
\hline
11. {\small MPF: KoralW }& ALEPH    & $0.0118\,(7)$ & $0.0103\,(9)$ 
& $0.0008\,(10)$ & CALO5  \\
12. {\small MPF: KoralW }& ALEPH    & $0.0122\,(7)$ & $0.0172\,(9)$ 
    & $0.0009\,(10)$ & CALO25 \\
\hline\hline
\end{tabular}
\end{center}
}
\caption{\small\sf
  The shifts on $\lambdag$ from various fits. 
  We use the abbreviations: Y=YFSWW3 and R=RacoonWW.
  The numbers in parentheses are the fit errors corresponding to
  the last digits of the results.
}
\label{tab:Tab1}
\end{table}

We are now ready to quantify various effects in the 
$\cos\theta_W^{\rm LAB}$ distribution 
in terms of $\lambdag$ using the MPF procedure.
To this end, we define
$\rho_{FD}(\cos\theta_W)=D_{1}(\cos\theta_W,0)/D_{2}(\cos\theta_W,0)$
where $D_{1}$ and $D_{2}$ are the normalized angular distributions
(for $\lambdag=0$), whose difference we want to quantify and fit this
distribution to the MPFF $\rho(\cos\theta_W,\lambdag)$ to obtain
$\Delta\lambdag$.
In Table~\ref{tab:Tab1},  columns 3--5, we show
$\Delta\lambdag$ due to the \Order{\alpha} EW NL corrections: in YFSWW3 
denoted by ``Y: Best$-$ISR'', 
in RacoonWW denoted by ``R: Best$-$ISR'',
and due to the difference RacoonWW$-$YFSWW3 in their ``Best'' modes, 
denoted by ``Best: R$-$Y''. 
In the first two rows of Table~\ref{tab:Tab1}, we include for the 
purpose of ``backward compatibility'' $\Delta\lambdag$
obtained by using the fitting procedure employing the FF of KorWan
(constructed originally for BARE$_{4\pi}$).
Rows 3--12 
in Table~\ref{tab:Tab1} show the results of the MPF procedure.
As already stressed, the type of the MPF is defined by the variant of the MC
and the type of the event selection, which are indicated in the first
and second columns.
In rows 3 and 4 of Table~\ref{tab:Tab1}, we use the MPFF
determined from the  MC run of ``Y-ISR'', 
i.e.\ the MPFF was constructed using the
results from YFSWW3 in the ``ISR'' mode (described above).
The similar MPFF from RacoonWW, denoted by ``R-ISR'' was used
to obtain the results in rows 5 and 6. 
In rows 7 and 8 ``Y-Best'' denotes the MPFF constructed from 
the run of YFSWW3 in the ``Best'' mode, and the similar one
for RacoonWW corresponds to rows 9 and 10 (``R-Best'').
The type of the event selection (CALO5 or CALO25)
in the MC data is indicated in the last column.
We use the MPFFs with the coefficients for CALO5 as it gives
practically the same results for 
the fitted $\lambdag$ as the ones for CALO25.
We see that all the results of the above five MPFs are
consistent with each other, 
and for YFSWW3 they also agree quite 
well with the results of the fits employing the FF of KorWan.

The results in Table~\ref{tab:Tab1} show that the NL corrections 
influence $\lambdag$ considerably; they shift $\lambdag$
by $-0.0096$ to $-0.0172$ with respect to the ISR-level predictions,
which is comparable to the expected
final experimental precision on this TGC at LEP2. 
The large differences in the shifts of $\lambdag$ for ``R: Best$-$ISR''
between CALO5 and CALO25 originate from the fact that the ``ISR''
prediction of RacoonWW includes only collinear photons and thus is not
influenced by photon recombination, in contrast to
the ``Best'' prediction of RacoonWW and the ``ISR'' and ``Best''
predictions of YFSWW.  The results of Table~\ref{tab:Tab1} demonstrate
that the NL corrections have to be included in the respective data analyses. 
On the other hand, the differences between the ``Best'' predictions
of RacoonWW and YFSWW3 are very small, $< 0.001$. This is important
as the two programs differ in many aspects of the implementation of various
effects. The good agreement between these programs is an
indication that the effects on $\lambdag$ due to 
the used approximations and the missing higher-order corrections 
should be small. This will be further investigated in the following.

Last but not least, the MPFF can also be defined for the true LEP2 acceptance.
In rows 11 and 12 we exploit MPFFs that use  $a_i$, $b_i$ and $c_i$
determined for the real-life ALEPH acceptance, which are denoted as ``ALEPH''.
The ALEPH selection and reconstruction efficiencies for jets and leptons are 
described in refs.~\cite{Aleph-MW:2000,Aleph-TGC:2001} 
(typically, the lepton acceptance reaches $|{\cos\theta}|$ = 0.95).
Again the values of the fitted $\Delta\lambdag$ are about the same as in
the previous exercises for the ``academic'' event selections.
This important cross-check
makes us confident that our estimates of the TU of $\lambdag$ are indeed 
relevant to the real-life LEP2 measurements.

\begin{table}[!ht]
{
\begin{center}
\begin{tabular}{||l|c|c||}
\hline\hline
  Effect  & Acceptance & $\Delta\lambdag$  \\
\hline\hline
 \raisebox{-1.5ex}[0cm][0cm]{1. Best $-$ ISR} & BARE$_{4\pi}$  
 & $0.0108\,(7)$  \\
 & CALO5$_{4\pi}$   
 & $0.0110\,(7)$  \\
\hline
 \raisebox{-1.5ex}[0cm][0cm]{2. ISR$_3$ $-$ ISR$_2$} & BARE$_{4\pi}$  
 & $0.0001\,(2)$  \\
 & CALO5$_{4\pi}$   
 & $0.0001\,(2)$  \\
\hline
 \raisebox{-1.5ex}[0cm][0cm]{3. FSR$_2$ $-$ FSR$_1$} & BARE$_{4\pi}$  
 & $0.0001\,(3)$  \\
 & CALO5$_{4\pi}$   
 & $0.0001\,(3)$  \\
\hline
 \raisebox{-1.5ex}[0cm][0cm]{4. $4f$-background corr. (Born)} & CALO5  
 & $0.0021\,(3)$  \\
 & CALO25   
 & $0.0021\,(3)$  \\
\hline
 \raisebox{-1.5ex}[0cm][0cm]{5. $4f$-background corr. (with ISR)} & CALO5  
 & $0.0005\,(3)$  \\
 & CALO25   
 & $0.0005\,(3)$  \\
\hline
 \raisebox{-1.5ex}[0cm][0cm]{6. EW-corr. scheme: $(B) - (A)$} & CALO5  
 & $0.0006\,(9)$  \\
 & CALO25   
 & $0.0006\,(9)$  \\
\hline
 \raisebox{-1.5ex}[0cm][0cm]{7. LPA$_b$ $-$ LPA$_a$} & CALO5  
 & $0.0017\,(9)$  \\
 & CALO25   
 & $0.0018\,(9)$  \\
\hline\hline
\end{tabular}
\end{center}
}
\caption{\small\sf
  The shifts on $\lambdag$ from various effects, obtained with YFSWW3
  and KoralW. 
  See the text for more details.
}
\label{tab:Tab2}
\end{table}

In Table~\ref{tab:Tab2}, we present the results of the 
$\lambdag$-shifts due to switching on/off various effects in 
YFSWW3 and KoralW.
They were obtained with the MPFF constructed with YFSWW3-Best for
CALO5. The subscript $4\pi$ in the acceptance denotes the full solid-angle 
coverage. 
In row 1, we show the results of the MPF for the NL correction
in YFSWW3, but for acceptances slightly different from the ones
in Table~\ref{tab:Tab1}, namely for BARE$_{4\pi}$ (no photon recombination) 
and CALO5$_{4\pi}$ (similar to CALO5, but without cuts 1 and 3).
These results agree very well with the corresponding
results in Table~\ref{tab:Tab1} for CALO5 and CALO25 (cf. column 3). 
This shows that the value of $\lambdag$ is not very sensitive 
to the choice of cuts and acceptances.
Rows 2 and 3 contain the results of switching off the 3rd order ISR
and the 2nd order FSR corrections, respectively. Both these effects
are negligible at LEP2 in terms of the $\lambdag$-shifts.
The $4f$-background corrections%
  \footnote{The complete Born-level $4f$ matrix element
            in KoralW was generated with the help of the GRACE2 
            package~\cite{GRACE2}.}
are investigated in rows 4 and 5. While a $4f$-background contribution 
at the Born level induces a shift on $\lambdag$ of $\sim 0.002$, 
it gives a negligible effect when it is combined with the ISR.
This is because the ISR affects the $\cos\theta_W^{\rm LAB}$ 
distribution in a way opposite to that of the $4f$-background,
which leads to large cancellations of the latter effect.
Then we investigate the uncertainties of the NL corrections due
to different schemes for the EW effective couplings (they account for
some parts of higher-order EW corrections): the so-called
schemes {\em (A)} and {\em (B)} in YFSWW3~\cite{yfsww3:2001} (row 6),
and due to the different LPA approaches: LPA$_a$ versus 
LPA$_b$~\cite{yfsww3:2001} (row 7).
While the effects on $\lambdag$ of the differences between schemes 
{\em (A)} and {\em (B)} are negligible (below the fit errors), 
the ones due to the LPA variation are $\sim 0.002$ -- this can be regarded 
as the LPA uncertainty in YFSWW3.

\begin{table}[!ht]
{
\begin{center}
\begin{tabular}{||l|c|c||}
\hline\hline
  Effect  & Acceptance & $\Delta\lambdag$  \\
\hline\hline
 \raisebox{-1.5ex}[0cm][0cm]{1. Best $-$ ISR} 
 & CALO5   & $0.0096\,(8)~\,$  \\
 & CALO25  & $0.0158\,(8)~\,$  \\
\hline
 \raisebox{-1.5ex}[0cm][0cm]{2. Off-shell Coulomb effect}
 & CALO5   & $0.0001\,(10)$  \\
 & CALO25  & $0.0001\,(10)$  \\
\hline
 \raisebox{-1.5ex}[0cm][0cm]{3. $4f$-background corr. (Born)} 
 & CALO5  & $0.0029\,(10)$  \\
 & CALO25 & $0.0029\,(10)$  \\
\hline
 \raisebox{-1.5ex}[0cm][0cm]{4. $4f$-background corr. (with ISR)} 
 & CALO5  & $0.0008\,(10)$  \\
 & CALO25 & $0.0008\,(10)$  \\
\hline
 \raisebox{-1.5ex}[0cm][0cm]{5. On-shell projection} 
 & CALO5  & $0.0003\,(10)$  \\
 & CALO25 & $0.0003\,(10)$  \\
\hline
 \raisebox{-1.5ex}[0cm][0cm]{6. DPA definition} 
 & CALO5  & $0.0005\,(10)$  \\
 & CALO25 & $0.0005\,(10)$  \\
\hline\hline
\end{tabular}
\end{center}
}
\caption{\small\sf
  The shifts on $\lambdag$ from various effects, obtained with RacoonWW. 
  See the text for more details.
}
\label{tab:Tab3}
\end{table}
In Table~\ref{tab:Tab3}, we show results of the $\lambdag$-shifts due
to switching on/off various effects in RacoonWW, as obtained with the
MPFF constructed with RacoonWW-Best for CALO5.  In row 1, we give the
shifts between the ``Best'' and ``ISR'' modes of RacoonWW. These
shifts are due to the NL electroweak corrections, including, in
contrast to the numbers from YFSWW3 in Table \ref{tab:Tab2}, also the
effects of non-collinear photon emission.  Row 2 exhibits the results
of switching off the off-shellness of the Coulomb singularity. This effect is
negligible, since the energy is not in the threshold region and the
Coulomb singularity affects mainly the normalization and thus cancels
in $D$ and $\rho$.  
The effect of the $4f$-background diagrams  without and with the
corresponding ISR (rows 3 and 4) is consistent
with the corresponding results in Table \ref{tab:Tab2}.  
Finally, we study the uncertainties
of the NL corrections due to different definitions of the DPA in
RacoonWW (rows 5 and 6) 
\cite{Denner:2000bj}.  These effects are
smaller than the corresponding effects seen in row 7 
of Table
\ref{tab:Tab2}. This is because, in RacoonWW, only the
virtual corrections are treated in DPA, while in YFSWW3 also the real
corrections feel the LPA.

\begin{table}[!ht]
{
\begin{center}
\begin{tabular}{||c|c||c|c|c|c||}
\hline\hline
\multicolumn{2}{||c||}{\bf ALEPH MPFF}  & 
\multicolumn{4}{|c||}{\bf Fitted data } \\
\hline
  Channel  & Acceptance & {Y: Best$-$ISR}  & {R: Best$-$ISR}  & 
  {Best: R$-$Y} & Acceptance\\
\hline\hline
\raisebox{-3.5ex}[0cm][0cm]{$\mu\nu_{\mu} qq$}  & 
\raisebox{-1.5ex}[0cm][0cm]{TRUE}  
    & $0.0118\,(7)$ & $0.0102\,(9)$ & $0.0008\,(10)$  & CALO5  \\
 &  & $0.0121\,(7)$ & $0.0170\,(9)$ & $0.0009\,(10)$  & CALO25 \\
\cline{2-6}
 & \raisebox{-1.5ex}[0cm][0cm]{RECO}  
    & $0.0118\,(7)$ & $0.0103\,(9)$ & $0.0008\,(10)$  & CALO5  \\
 &  & $0.0122\,(7)$ & $0.0172\,(9)$ & $0.0009\,(10)$  & CALO25 \\
\hline\hline
\raisebox{-3.5ex}[0cm][0cm]{$e\nu_{e} qq$}  & 
\raisebox{-1.5ex}[0cm][0cm]{TRUE}  
    & $0.0119\,(7)$ & $0.0103\,(9)$ & $0.0008\,(10)$  & CALO5  \\
 &  & $0.0122\,(7)$ & $0.0172\,(9)$ & $0.0009\,(10)$  & CALO25 \\
\cline{2-6}
 & \raisebox{-1.5ex}[0cm][0cm]{RECO}  
    & $0.0119\,(7)$ & $0.0103\,(9)$ & $0.0008\,(10)$  & CALO5  \\
 &  & $0.0123\,(7)$ & $0.0172\,(9)$ & $0.0009\,(10)$  & CALO25 \\
\hline\hline
\raisebox{-3.5ex}[0cm][0cm]{$\tau\nu_{\tau} qq$}  & 
\raisebox{-1.5ex}[0cm][0cm]{TRUE}  
    & $0.0115\,(7)$ & $0.0100\,(8)$ & $0.0008\,(10)$  & CALO5  \\
 &  & $0.0118\,(7)$ & $0.0166\,(8)$ & $0.0009\,(10)$  & CALO25 \\
\cline{2-6}
 & \raisebox{-1.5ex}[0cm][0cm]{RECO}  
    & $0.0107\,(6)$ & $0.0091\,(8)$ & $0.0007\,(9)~\,$  &
    {CALO5$_{\rm RECO}$} \\
 &  & $0.0109\,(6)$ & $0.0152\,(8)$ & $0.0008\,(9)~\,$  &
    {CALO25$_{\rm RECO}$}\\
\hline\hline
\raisebox{-3.5ex}[0cm][0cm]{$qqqq$}  & 
\raisebox{-1.5ex}[0cm][0cm]{TRUE}  
    & $0.0118\,(7)$ & $0.0102\,(9)$ & $0.0008\,(10)$  & CALO5  \\
 &  & $0.0120\,(7)$ & $0.0169\,(9)$ & $0.0009\,(10)$  & CALO25 \\
\cline{2-6}
 & \raisebox{-1.5ex}[0cm][0cm]{RECO}  
    & $0.0094\,(6)$ & $0.0081\,(7)$ & $0.0007\,(8)~\,$  & 
 {CALO5$_{\rm RECO}$} \\
 &  & $0.0096\,(6)$ & $0.0132\,(7)$ & $0.0008\,(8)~\,$  &
 {CALO25$_{\rm RECO}$}\\
\hline\hline
\end{tabular}
\end{center}
}
\caption{\small\sf
  The results for $\lambdag$ shifts for the fits using the   
  ALEPH fitting functions obtained at the parton level (TRUE) and 
  with the full detector reconstruction (RECO) for various channels. 
  The subscript RECO for some CALO5 and CALO25 acceptances means that the
  reconstruction effects were also included in the corresponding MC data
  (see the text for details).
  The fitted data are always for the channel $\mu^-\bar{\nu}_{\mu} u\bar{d}$.
  We use the abbreviations: Y=YFSWW3 and R=RacoonWW. 
}
\label{tab:Tab4}
\end{table}

In order to see whether our results can be extended also to other channels, 
we performed the following exercise. Using simulated data at the parton
level (denoted as TRUE) and with the full ALEPH detector reconstruction
(denoted as RECO) for the channels $\mu\nu_{\mu} qq$, 
$e\nu_{e} qq$, $\tau\nu_{\tau} qq$ and $qqqq$, we constructed eight MPFFs 
-- two MPFFs per channel. Then we fitted these functions to our MC data,
the same as in Table~\ref{tab:Tab1}, for the channel 
$\mu^-\bar{\nu}_{\mu} u\bar{d}$.
The results of these fits for all the TRUE-level fitting functions 
and for the $\mu\nu_{\mu} qq$ and $e\nu_{e} qq$ RECO-level ones
are shown in Table~\ref{tab:Tab4}.
As one can see, all the TRUE-level results are
consistent with each other and agree with the ones for the YFSWW3 
and RacoonWW MPFFs (cf. Table~\ref{tab:Tab1}, rows 3--10). 
The RECO-level results agree very well
with the TRUE-level ones for the channels $\mu\nu_{\mu} qq$ and $e\nu_{e} qq$.
The above results show that the experimentally reconstructed distributions
of $\cos\theta_W$ for the channels  $\mu\nu_{\mu} qq$ and $e\nu_{e} qq$
are very similar to the parton-level ones and also to each other,
which in our exercise resulted in almost identical fitting functions. 
This leads us to the conclusion that our previous findings for the
channel $\mu\nu_{\mu} qq$ can be extended to the channel
$e\nu_{e} qq$.

For the $\tau\nu_{\tau} qq$ and $qqqq$ channels, the reconstructed
distributions differ considerably from the TRUE-level ones.  We have
checked that fitting the RECO-level MPFF from the $qqqq$
($\tau\nu_{\tau} qq$) channel to our MC data results in the shifts of
$\lambdag$ which are by $\sim 100\%$ ($\sim 25\%$) larger than the
corresponding ones for the TRUE-level MPFF (these results are not
shown in Table \ref{tab:Tab4}).
Thus, the parton-level MC data are not appropriate for estimating
the shifts of $\lambdag$ in the real-life experimental measurements
for these channels.
In order to obtain realistic effects, these data have to be processed
through the full detector simulation. Instead of feeding our MC events
into the full ALEPH reconstruction generator, we applied so-called transfer 
matrices to the MC $\cos\theta_W^{\rm LAB}$ distributions.
A transfer matrix gives the probability for an event generated in
a TRUE $\cos\theta_W^{\rm LAB}$ bin to be found in a given RECO bin.
This takes into account non-diagonal terms induced by the ALEPH
detector effects
(jet resolution, jet pairing, jet charge, and $\tau$ reconstruction)
\cite{Aleph-TGC:2001}.
The appropriate transfer matrices for these channels were constructed 
using the Monte Carlo samples generated within the full ALEPH detector
simulation environment~\cite{Aleph-Detector:1994}.
Then, we used such experimental-like distributions in the MPFs 
with the RECO-level fitting functions. In these fits we used
appropriate error (covariance) matrices, taking into account correlations
between the bins and between the transfer-matrix elements.
The results for these two channels at the RECO-level 
complete Table~\ref{tab:Tab4}
(they are denoted by the subscript RECO attached to the respective 
acceptance in the last column). 
As can be seen, the values of the $\lambdag$ shifts are close to the
corresponding TRUE-level results. We also checked that this method 
did not affect the $\mu\nu_{\mu} qq$ and $e\nu_{e} qq$ channels, as expected.
Therefore our conclusions for the TU of $\lambdag$ in 
the $\mu^-\bar{\nu}_{\mu} u\bar{d}$ channel can be applied also
to the $\tau\nu_{\tau} qq$ and $qqqq$ channels in the realistic
experimental analyses.

From the above numerical exercises and the accompanying discussion, 
we come to the following conclusions:
\begin{itemize} 
\item 
  The non-leading EW corrections cause shifts of 
  $\lambda=\lambda_\gamma=\lambda_Z$ at the level of $0.01$--$0.02$, 
  which is comparable to the combined experimental accuracy 
  for this anomalous TGC at LEP2.
  Thus, these corrections have to be taken into account in the
  experimental analyses.

\item The comparisons between YFSWW3 and RacoonWW 
  and the tests of various modes/options in both programs
  (cf.\ Tables~\ref{tab:Tab2} and \ref{tab:Tab3}) allow us to estimate
  the EW theoretical uncertainty in $\lambdag$. We use the largest
  shifts found in these comparisons and apply a safety
  factor of $2$ to account for the $\sim 30\%$ sensitivity 
  loss due to the single-distribution fit and for possible higher-order
  effects missing in both programs. 

  From this we estimate the EW theoretical uncertainty in $\lambdag$ 
  of the MC tandem KoralW\&YFSWW3 and of the MC program RacoonWW
  to be $\sim 0.005$ at LEP2 energies,
  which is $\sim 1/2$ of the expected combined experimental error. 

\end{itemize}

\newpage
\noindent
{\bf Acknowledgments}

S.J., W.P., M.S., B.F.L.W. and Z.W.
warmly acknowledge the kind support of the CERN TH and EP divisions.
One of us (B.F.L.W.) thanks Profs.\ S.~Bethke and L.~Stodolsky for the
kind hospitality and support of the Werner-Heisenberg-Institut, MPI, Munich,
and thanks Prof. C. Prescott for the kind hospitality of SLAC Group A.


\end{document}